\documentclass[prl,twocolumn,showpacs]{revtex4}

\usepackage{graphicx}
\usepackage{dcolumn}
\usepackage{bm}

\begin{document}

\preprint{FZJ-IKP}

\title{Measurement of the Analyzing Power in $\vec{p}d \to
  (pp)n$ \\with a Fast Forward $^1S_0$--Diproton}

\author{
S.~Yaschenko$^{1,2}$,
S.~Dymov$^{2}$,
A.~Kacharava$^{1,3}$,
V.~Komarov$^{2}$,
G.~Macharashvili$^{2,3}$,
F.~Rathmann$^{4}$\footnote{Electronic address: f.rathmann@fz-juelich.de},
S.~Barsov$^{5}$,
R.~Gebel$^{4}$,
M.~Hartmann$^{4}$,
A.~Khoukaz$^{6}$,
P.~Kulessa$^{4,7}$,
A.~Kulikov$^{2}$,
V.~Kurbatov$^{2}$,
N.~Lang$^{6}$,
I.~Lehmann$^{4}$,
B.~Lorentz$^{4}$,
T.~Mersmann$^6$,
S.~Merzliakov$^{2}$,
S.~Mikirtytchiants$^{5}$,
A.~Mussgiller$^{4}$,
M.~Nioradze$^{3}$,
H.~Ohm$^{4}$,
D.~Prasuhn$^{4}$,
R.~Schleichert$^{4}$,
H.~Seyfarth$^{4}$,
E.~Steffens$^1$
H.~J.~Stein$^{4}$,
H.~Str\"oher$^{4}$,
Yu.~Uzikov$^{2,8}$,
B.~Zalikhanov$^{2}$, and
N.~Zhuravlev$^{2}$
}

\affiliation{$^1$Physikalisches Institut II, Universit\"at
  Erlangen-N\"urnberg, 91058 Erlangen, Germany}

\affiliation{$^2$Laboratory of Nuclear Problems, Joint Institute for
  Nuclear Research, 141980 Dubna, Russia}

\affiliation{$^3$High Energy Physics Institute, Tbilisi State
  University, 380086 Tbilisi, Georgia}

\affiliation{$^4$Institut f\"ur Kernphysik, Forschungszentrum
  J\"ulich, 52425 J\"ulich, Germany}

\affiliation{$^5$High Energy Physics Department, Petersburg Nuclear
  Physics Institute, 188350 Gatchina, Russia}

\affiliation{$^6$Institut f\"ur Kernphysik, Universit\"at M\"unster,
  48149 M\"unster, Germany}

\affiliation{$^7$Henryk Niewodniczanski Institute of Nuclear Physics,
  31342, Cracow, Poland}
       
\affiliation{$^8$Kazakh National University, Department of Physics,
  480012 Almaty, Kazakhstan}

\date{\today}

\begin{abstract}
  A measurement of the analyzing power $A_y$ of the $\vec{p}d \to (pp)
  + n$ reaction was carried out at the ANKE spectrometer at COSY at
  beam energies of 0.5 and 0.8~GeV by detection of a fast forward
  proton pair of small excitation energy $E_{pp} < 3$~MeV.  The
  $S$--wave dominance in the fast diproton is experimentally
  demonstrated in this reaction.  While at $T_p=0.8$~GeV the measured
  analyzing power $A_y$ vanishes, it reaches almost unity at
  $T_p=0.5$~GeV for neutrons emitted at $\theta_n^{c.m.}=167^\circ$.
  The results are compared with a model taking into account
  one--nucleon exchange, single scattering, and $\Delta$ (1232)
  excitation in the intermediate state. The model describes fairly
  well the unpolarized cross section obtained earlier and the
  analyzing power at 0.8~GeV, \ it fails to reproduce $A_y$ at
  0.5~GeV.

\end{abstract}

\maketitle

The structure of the lightest nuclei at short distances ($r_{NN} <
0.5$~fm) or high relative momenta ($q > 1/r_{NN}\sim 0.4$~ GeV/c), and
the closely related nucleon--nucleon ($NN$) interaction constitute
fundamental problems in nuclear physics. Experimental investigations
employ processes where the momentum transfer to the nucleus is large
($Q \sim 1$~GeV/c). Most of our present knowledge about the structure
of the deuteron has been obtained from electromagnetic probes. The
existing data on elastic electromagnetic deuteron form factors for $Q
< 1$~GeV/c are in reasonable agreement with $NN$ models based on the
exchange of mesons \cite{garcon,gilman}.  The situation above $Q\sim
1$~GeV/c becomes much less clear due to increasing contributions from
meson--exchange currents (MEC) in $ed$ interactions and theoretical
uncertainties in their treatment.  Moreover, meson--exchange models
have difficulties to explain photo--disintegration data ($\gamma\,d\to
np$) for energies $E_\gamma > 1$~GeV \cite{bochna}.  Models based on
quark degrees of freedom have recently become quite successful in
describing the data \cite{grishina}.

Independent information about the short--range structure of nuclei can
be obtained from hadronic interactions at large $q$. However, the
study of the simplest processes in the GeV region, $pd\to dp$
\cite{ashgirey} as well as inclusive ($dp\to pX$ \cite{dubna}) and
exclusive ($pd\to ppn$ \cite{belost}) deuteron disintegration turned
out to be not conclusive in this respect \cite{garcon}.  It is
therefore important to obtain new data under conditions that make the
theoretical interpretation more transparent. Recently, the unpolarized
cross section of the $pd\to(pp)n$ reaction was measured at proton beam
energies $T_p=0.6$ to 1.9~GeV in a kinematics similar to backward $pd$
elastic scattering \cite{komarov} with formation of a fast diproton in
a $^1S_0$ state at low excitation energy ($E_{pp} < 3$~MeV).  At high
$Q$, near threshold deuteron electro--disintegration $d(e,e')pn$
\cite{auffert} and single pion production, $pp\to pp\pi^0$
\cite{meyer} and $pn\to(pp)\pi^-$ \cite{duncan}, constitute prominent
examples for the observation that substitution of an ordinary deuteron
in the final state by a singlet deuteron or its isotriplet partner,
the diproton, will give new insight into the reaction dynamics.  In
$pd\to(pp) n$, the diproton provides two new features which are absent
for isosinglet nucleon pairs \cite{ashgirey,dubna,belost}.  {\it i)}
The contribution from three--body forces, related to two isovector
meson exchanges, in particular the excitation of $\Delta$ and $N^*$
resonances in the intermediate state, is suppressed by the isospin
factor of 3 in the amplitude of this process \cite{yujetp}.  This
suppression is of relevance, because the theoretical interpretation of
three--body effects in hadronic reactions encountered problems similar
to those of MEC in electromagnetic processes
\cite{kondratyuk,budardillig,yujetp,sakai}.  {\it ii)} The $S$--wave
dominates in the internal state of the diproton at $E_{pp} < 3$~MeV.
Due to the repulsive $NN$ core, the $^1S_0$ diproton wave function
$\psi_{pp}(q)$ has a node at relative $pp$ momenta $q\approx
0.4$~GeV/c \cite{imusmuz}, leading to a distinct energy dependence of
various polarization observables \cite{yujpg} that helps to identify
the dominant reaction mechanisms.  The recent analysis
\cite{haidenbauer} of the $pd \to (pp)_{^1S_0}n$ cross section, based
on a model for the $pd\to dp$ process \cite{kondratyuk}, includes the
one--nucleon exchange (ONE), single scattering (SS), and double $pN$
scattering with excitation of a $\Delta(1232)$ isobar.  This analysis
accounts for initial and final state interactions and employs modern
$NN$ potentials, based on the exchange of mesons, e.g. CD--Bonn
\cite{machleidt}. A reasonable agreement with our recent data
\cite{komarov} is achieved.  In contrast, the widely used $NN$
potentials like the Paris \cite{paris} and especially the Reid Soft
Core (RSC) potential \cite{rsc} lead to a strong disagreement with the
data. These potentials apparently overestimate the high--momentum
components of the $NN$ wave function $\psi_{NN}(q)$.  Another
approach, using the ONE mechanism only \cite{kaptari}, which is based
on relativistic $P$--waves in the diproton and deuteron also explains
the data \cite{komarov}.
 
New information about this reaction can be obtained from measurements
of polarization observables. Here we report about the first
measurement of the vector analyzing power $A_y$ at $T_p=0.5$ and
0.8~GeV of the reaction \begin{equation} \vec{p} + d \to (pp)_{^1S_0} + n\,,
\end{equation} where $(pp)_{^1S_0}$ denotes a fast proton
pair emitted in the forward direction with small excitation energy
$E_{pp} < 3$~MeV. The two beam energies were chosen because of the
difference in the reaction mechanisms predicted by the model
\cite{haidenbauer}. While at 0.5~GeV the contribution from the
$\Delta$ excitation is comparable to that from ONE, at 0.8~GeV the
latter is completely eliminated due to the node in the $pp$ wave
function $\psi_{pp}(q)$ and hence the process is governed by the
$\Delta$ mechanism. Each mechanism under consideration alone yields an
almost vanishing analyzing power. Because of their interference a
substantial $A_y$ arises, which is expected to decrease with
increasing beam energy between $T_p=0.5$ and 0.8~GeV.

\begin{figure}[h]
\includegraphics[width=7.9cm]{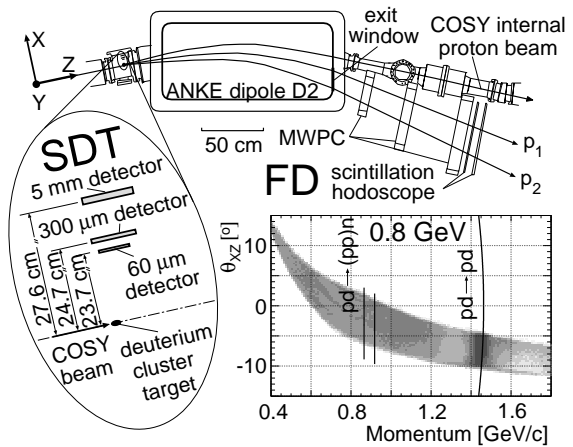}
\caption{\label{fig:setup} 
  Top--view of the ANKE spectrometer with the forward detector (FD)
  and the Silicon--detector telescope (SDT, see inset).  Diprotons
  from the breakup reaction stem from the region indicated in the
  distribution of the polar angle projection $\theta_{xz}$ vs particle
  momentum (lower right).  Protons from $pd$ elastic scattering are
  distributed along the kinematical locus.  }
\end{figure}

The experiment was performed at the ANKE spectrometer \cite{barsov} at
the internal beam of COSY--J\"ulich \cite{maier} with about $3 \cdot
10^{9}$ stored vertically polarized protons. The experimental setup is
shown in Fig.~\ref{fig:setup}.
The Forward Detector (FD) measured proton pairs from the deuteron
breakup and single protons, scattered at small angles from $pd \to
pX$.  The Silicon--detector telescope (SDT) recorded recoil deuterons
from small--angle elastic $pd$ scattering. The FD
\cite{chiladze,dymov} comprises a set of three multi--wire
proportional chambers (MWPCs) and a two--plane scintillation
hodoscope, consisting of vertically oriented counters (8 in the first
plane, 9 in the second).  The acceptance of the setup
(Fig.~\ref{fig:setup}) allows one to detect protons from
quasi--elastic scattering from $\theta_{\rm lab}=4.5^{\circ}$ to
$10^{\circ}$ at $T_p=0.5$~GeV and 0.8 GeV. The vertical acceptance
corresponds to $\pm 3.5^{\circ}$.  Protons from the breakup reaction
with $E_{pp}<3$ MeV are detected from $\theta_{\rm lab}=0^{\circ}$ to
6.5$^{\circ}$ at both energies, the polar angles of the proton pairs
range from $\theta^{\rm c.m.}_{pp}=0^{\circ}$ to 14$^{\circ}$.  The
uncertainty in $E_{pp}$ ranges from 0.2~MeV at $E_{pp}$ = 0.3~MeV to
0.3~MeV at $E_{pp}$ = 3 MeV.  The SDT \cite{lehmann} consists of three
layers of silicon counters in the horizontal plane located inside the
vacuum of the ANKE target chamber. Recoil deuterons at angles around
$\theta_{\rm lab}=90^{\circ}$ were detected in the SDT in coincidence
with elastically scattered protons in the FD.  The SDT provided an
unambiguous deuteron identification with a detector resolution of
300~keV.  The deuterium cluster--jet \cite{khoukaz} produced a target
density of about $2 \cdot 10^{14}$~atoms/cm$^{2}$ with a target length
along the beam of 12~mm and a width of 4.9~mm.

The tracks were reconstructed from the hits in the MWPCs, ensuring
that they intercept the 0.5~mm Al exit window.  The three--momentum
vectors were determined by tracing the particles through the magnetic
field of the spectrometer \cite{dymov}. For two particles hitting
different hodoscope counters the correlation of the measured
time--of--flight (TOF) difference $\Delta t_{\rm meas}$ and $\Delta
t(\vec{p_1},\vec{p_2})$,
\begin{figure}[hbt]
\includegraphics[width=6.4cm]{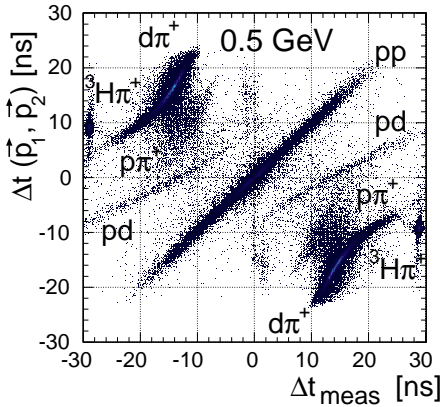}
\caption{\label{fig:pid} Proton pairs are identified from the correlation 
  of the TOF differences $\Delta t_{\rm meas}$ and $\Delta
  t(\vec{p_1},\vec{p_2})$.}
\end{figure} 
calculated from the measured three--momenta assuming proton masses,
allows one to identify charged particle pairs from different reactions
(Fig.~\ref{fig:pid}).  However, proton pairs from the deuteron breakup
can be identified via missing mass without this TOF criterion, as
discussed in Ref.~\cite{komarov}. At both energies and for both
orientations of the beam polarization, the missing mass peak is
observed at the neutron mass $M_n$, yielding $(0.938 \pm 0.005)$
$\mathrm{GeV/c^2}$ ($T_p=0.5$~GeV) and $(0.935 \pm 0.005)$
$\mathrm{GeV/c^2}$ ($T_p=0.8$~GeV).  The (rms) peak widths are
16~$\mathrm{MeV/c^2}$ and $\mathrm{20~MeV/c^2}$, respectively.

\begin{figure}[hbt]
\includegraphics[width=7.9cm]{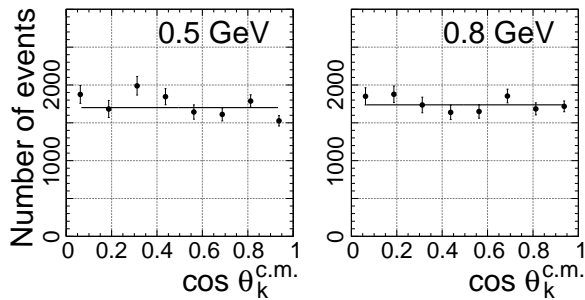}
\caption{\label{fig:swave} 
  Acceptance corrected distribution of events as function of
  $\cos\theta_k^{c.m.}$ for 0.5~GeV (left panel) and 0.8~GeV (right).}
\end{figure}

The $S$--wave dominance in the diproton final state is illustrated in
Fig.~\ref{fig:swave}, where the acceptance corrected distribution of
events is shown over the cosine of the proton polar angle
($\cos\theta_k^{c.m.}$) in the two--proton rest frame with respect to
the total momentum of the pair.

The spin--dependent yield is given by
\begin{eqnarray}
Y(\theta,\phi)_{\uparrow(\downarrow)}= Y_0(\theta)
\cdot [1\ + \ P_{\uparrow(\downarrow)} \cdot A_y(\theta) \cdot \cos\phi],
\end{eqnarray}
where $Y_0$ denotes the spin--averaged yield and $P_{\uparrow
  (\downarrow)}$ the absolute value of the beam polarization, oriented
along the vertical $y$--axis. (The coordinate system is shown in
Fig.~\ref{fig:setup}.)  Polar and azimuthal angles $\theta$ and $\phi$
of the breakup reaction are determined from the neutron momentum in
the c.m.\ system, $\vec p_n = - (\vec p_{1}+ \vec p_{2})$, where $\vec
p_{1}$ and $\vec p_{2}$ are the proton momenta.

The absence of azimuthal symmetry of the ANKE spectrometer does not
permit one to measure a vector analyzing power from the left--right
count rate asymmetry. Therefore, we measured the analyzing power by
reversing every two cycles the orientation of the polarization.
Careful monitoring of the relative luminosity $L_{\uparrow}
/L_{\downarrow}$ was achieved by either detecting single particles in
the FD at $\theta_{\rm lab} < 1^{\circ}$ or at
$\phi=90^\circ\pm5^\circ$ and $\phi=270^\circ\pm5^\circ$, where the
rates are insensitive to the vertical beam polarization.

The beam polarization at $T_p=0.800$~GeV was determined from precise
$pd$--elastic analyzing power data \cite{irom} at 0.796~GeV.  The $pd$
elastic scattering angles were determined from the energy deposit of
the identified deuterons in the SDT. Since there are no data available
at 0.5 GeV, we resorted to the polarization export technique
\cite{pollock} to obtain a calibrated polarization for 0.5~GeV.  This
was achieved by setting up a cycle with a flat top at energy
$T_p=0.8$~GeV (I), followed by deceleration to a flat top at 0.5~GeV
(II), and subsequent re--acceleration to a flat top at 0.8~GeV (III).
Avoiding depolarization during crossing of the resonances, the
measured beam polarizations $P_I=0.564 \pm 0.003^{\mathrm{stat.}} \pm
0.004^{\mathrm{syst.}}$ and $P_{III} = 0.568 \pm
0.004^{\mathrm{stat.}} \pm 0.005^{\mathrm{syst.}}$ agree within
errors.  The systematic errors arise from the statistical uncertainty
of the relative luminosity.  The weighted average of $P_I$ and
$P_{III}$ was used to export the beam polarization to flat top II and
to determine the angular distribution of the previously unknown
analyzing power of $pd$ elastic scattering at 0.5 GeV. A small
angle--independent correction of $-0.0024$ was applied in the export
procedure to account for the 4~MeV difference in beam energy, using
the energy dependence of $A_y$ between 500 and 800 MeV.

\begin{figure}[hbt]
\includegraphics[width=7.9cm]{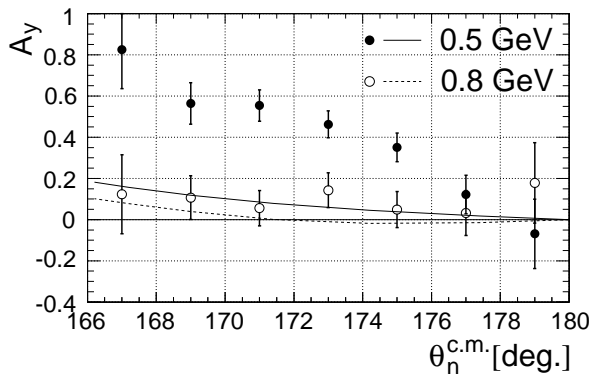}
\caption{\label{fig:ay} Angular dependence of the analyzing power $A_y$
  as function of the neutron polar angle $\theta_n^{c.m.}$ for
  $T_p=0.5$ ($\bullet$) and 0.8~GeV ($\circ$). The lines show
  predictions for $A_y$ at 0.5~GeV (solid) and 0.8~GeV (dashed) from
  the ONE + SS +$\Delta$ model~\cite{yujpg,haidenbauer}, with the
  CD--Bonn potential.}
\end{figure}

The analyzing power is determined from
\begin{equation} 
A_y(\theta)=\frac{\varepsilon(\theta)}{P} \cdot 
\frac{1}{\langle \cos\phi \rangle}_\theta\;,
\end{equation}
where $P=(P_\uparrow + P_\downarrow)/2$ and $\varepsilon(\theta)$ is
given by
\begin{equation} 
\varepsilon(\theta)=
\frac{N_{\uparrow}(\theta)/L_{\uparrow}-N_{\downarrow}(\theta)/L_{\downarrow}}
{N_{\uparrow}(\theta)/L_{\uparrow}+N_{\downarrow}(\theta)/L_{\downarrow}}\;.
\end{equation}
Here $N_{\uparrow}(\theta)/L_{\uparrow}$ and
$N_{\downarrow}(\theta)/L_{\downarrow}$ denote the number of events in
each $\theta$ bin, weighted by the relative luminosity for each
orientation of the beam polarization. Events were selected for which
$|\phi| \le 45^\circ$. The average $\langle \cos\phi \rangle_\theta =
N_\theta^{-1} \sum_{i} ^{N_\theta} (\cos\phi_{i})_{\theta}$, where
$N_\theta=N_\uparrow(\theta)+N_\downarrow(\theta)$, is determined from
the experimental data for each $\theta$ bin.  The number of counts
$N_{\uparrow}$ and $N_{\downarrow}$ were obtained from the neutron
missing mass spectra for proton pairs with $E_{pp}<3$ MeV. The spectra
for the two orientations of the beam polarization were fitted
separately with the sum of a Gaussian and a linear function to account
for the background and the yield was determined within a $\pm3\sigma$
range around $M_n$.  The background was subtracted separately for each
reconstructed missing mass value.  The obtained values of $A_y$ at 0.5
and 0.8~GeV are shown in Fig.\ \ref{fig:ay} as function of
$\theta_n^{c.m.}$ \cite{epaps}.

The systematic uncertainty of the analyzing power contains
contributions from various sources, which were all added in
quadrature. An upper limit for the difference of the beam polarization
$\Delta P= (P_\uparrow -P_\downarrow)/2=0.013$ was determined from a
polarization measurement using the low energy polarimeter of COSY. The
analyzing powers change by a factor $(1+\Delta P \cdot A_y)^{-1}$,
thus leading to a systematic error of at most $\pm0.008$.  The
systematic effect on $A_y$ due to the uncertainty of the relative
luminosity $L_\uparrow/L_\downarrow$ does not exceed $\pm 0.003$. The
total systematic uncertainty of $A_y$ is smaller than 20\% of the
statistical error and never exceeds $\pm0.02$ at all angles.
Finite--bin corrections to the final $A_y$ amount to at most 0.017,
nevertheless they were applied in all $\theta$ bins.

The measured $A_y$ is almost zero at 0.8~GeV, in agreement with the
predictions of the ONE+SS+$\Delta$ model. At this energy, the
calculated $A_y$ is almost insensitive to the spin structure of the
$\Delta$--mechanism, which completely dominates the process and
produces alone a near zero value of $A_y$.  A peculiarity of the data
at 0.5 GeV is the rapid increase of $A_y$ up to a value of 0.8 in a
small angular interval from 180$^\circ$ to 167$^{\circ}$.  (In
contrast, in the same angular interval the analyzing power of $pd \to
dp$ scattering \cite{p21, p22, p23} at comparable energies, 0.425,
0.68, 0.8 and 1.053~GeV, never exceeds 0.12.) The increase of $A_y$
with decreasing energy from 0.8 to 0.5~GeV is expected from the
ONE+SS+$\Delta$ model, however, the magnitude is grossly
underestimated. Different $NN$--interaction potentials (RSC, Paris) do
not improve the description. We recall that the model describes
reasonably well the unpolarized cross section \cite{haidenbauer}
($\chi^2/\mathrm{d.f.}=1.8$) and has no free parameters. The
inadequateness of the model to describe $A_y$ at 0.5~GeV may be
related to the spin structure of the $\Delta$--mechanism, given here
by the Born term of the $\pi+\rho$ meson exchange in the $NN\to \Delta
N$ transition \cite{kondratyuk,haidenbauer}.  (A similar problem was
observed recently \cite{sakai} at lower energies, where, however, the
role of the $\Delta$ is much weaker.) Improvements may be achieved by
using a coupled channel $NN$ and $N\Delta$ calculation as was done for
the $pN\to(pp)_{^1S_0}\pi$ process \cite{duncan}.

In summary, a large analyzing power is observed in the $\vec{p}d \to
(pp)_{^1S_0}n$ process at 0.5 GeV and a value close to zero at 0.8
GeV, significantly differing from the behavior of $A_y$ in $pd$
backward elastic scattering. The observed disagreement of the
ONE+SS+$\Delta$ model predictions with the measured $A_y$ clearly
demonstrates the need to reconsider the spin structure of three--body
forces related to the $\Delta$--mechanism. Further insight into the
short--range structure of the deuteron can be achieved from a
measurement of the tensor analyzing power $T_{20}$ in $p\vec{d} \to
(pp)_{^1S_0}n$, in preparation at ANKE, for which the theoretical
predictions are more robust than for $A_y$.

We acknowledge the support of the COSY accelerator crew, the help of
H.  Rohdje{\ss} (EDDA Collaboration) during the first beam
polarization measurements, and the temporary appointment of one of us
(I.L.) by FZ--Rossendorf.  This work was supported by the BMBF (06
ER126), a BMBF grant to JINR, the BMBF/WTZ grants (Rus--667--97, Rus
00/211, Rus 01/691, Kaz 99/001, and Kaz 02/001) and the
Heisenberg--Landau program.


\begin{thebibliography}{99}
\bibitem{garcon} M. Gar\c{c}on and J.W. Van Orden, Adv. Nucl. Phys. {\bf    26}, 293 (2001).
\bibitem{gilman} R. Gilman and F. Gross, J. Phys. {\bf G28}, R37  (2002).
\bibitem{bochna} C. Bochna {\it et al.}, Phys. Rev. Lett. {\bf 81},
  4576 (1998), see also references [3], [4], [9], and [10] therein.
\bibitem{grishina} V.Yu.~Grishina {\it et al.}, Eur. Phys. J. {\bf  A10}, 355 (2001) and 
{\bf A19}, 117 (2004).
\bibitem{ashgirey}L.S. Azhgirey {\it et al.}, Phys. Lett. {\bf B391}, 22 (1997).
\bibitem{dubna} A. Ableev {\it et al.}, Nucl. Phys. {\bf A393}, 491  (1983).
\bibitem{belost} S.L. Belostotsky {\it et al.}, Phys. Rev. {\bf C56},  50 (1997).
\bibitem{komarov} V. Komarov {\it et al.}, Phys. Lett. {\bf B553},  179 (2003).
\bibitem{auffert} S. Auffert {\it et al.}, Phys. Rev. Lett. {\bf 55}, 1362 (1985).
\bibitem{meyer}H.O. Meyer {\it et al.}, Nucl. Phys. {\bf A539}, 633 (1992).
\bibitem{duncan} F. Duncan {\it et al.}, Phys. Rev. Lett. {\bf 80}, 4390 (1998).
\bibitem{yujetp} Yu.N. Uzikov, JETP Lett. {\bf 75}, 5 (2002).
\bibitem{kondratyuk} L.A. Kondratyuk {\it et al.}, Phys. Lett. {\bf B100}, 448 (1981).
\bibitem{budardillig} A.~Boudard and M.~Dillig, Phys. Rev. {\bf C31},  302 (1985).
\bibitem{sakai} H. Sakai {\it et al.}, Phys. Rev. Lett. {\bf 84}, 5288  (2000).
\bibitem{imusmuz} O. Imambekov, Yu.N. Uzikov, Sov. J. Nucl. Phys. {\bf
    47}, 862 (1990). A.V. Smirnov and Yu.N. Uzikov, Phys. Atom. Nucl.  {\bf 61}, 361 (1998).
\bibitem{yujpg} Yu.N. Uzikov J. Phys. {\bf G28}, B13 (2002).
\bibitem{haidenbauer} J. Haidenbauer and Yu.N. Uzikov, Phys. Lett.  {\bf B562}, 227 (2003).
\bibitem{machleidt} R. Machleidt, Phys. Rev. {\bf C63}, 024001  (2001).
\bibitem{paris}M.  Lacombe {\it et al.}, Phys. Lett. {\bf B101}, 139  (1981).
\bibitem{rsc}J.R.V.  Reid, Ann. Phys. (N.Y.) {\bf 50}, 411 (1968).
\bibitem{kaptari} L.P. Kaptari {\it et al.}, Eur. Phys. J. {\bf A19},  301 (2004).
\bibitem{barsov} S. Barsov {\it et al.}, Nucl. Instrum. Methods {\bf    A462}, 364 (2001).
\bibitem{maier} R. Maier, Nucl. Instrum. Methods {\bf A390}, 1 (1997).
\bibitem{chiladze} B. Chiladze {\it et al.}, Part.  Nucl., Lett. {\bf    4}, 95 (2002).
\bibitem{dymov} S. Dymov {\it et al.}, Part. Nucl., Lett. {\bf 2}, 40  (2004).
\bibitem{lehmann} I. Lehmann {\it et al.}, Nucl. Instrum. Methods {\bf    A530}, 275 (2004).
\bibitem{khoukaz} A. Khoukaz {\it et al.}, Eur. Phys. J. {\bf D5}, 275  (1999).
\bibitem{irom} F. Irom {\it et al.}, Phys. Rev. {\bf C28}, 2380  (1983).
\bibitem{pollock} R.E. Pollock {\it et al.}, Phys. Rev. {\bf E55},  7606 (1997).
\bibitem{epaps}See EPAPS document No. ??? for a table of the results.
\bibitem{p21} N.E. Booth {\it et al.}, Phys. Rev. {\bf D4}, 1261  (1971).
\bibitem{p22} E. Biegert {\it et al.}, Phys. Rev. Lett. {\bf 41}, 1098 (1978).
\bibitem{p23} E. Winkelmann {\it et al.}, Phys. Rev. {\bf C21}, 2535 (1980).
\end{thebibliography}
\end{document}